\documentclass[letter,traditabstract]{aa}

\usepackage{graphicx,txfonts,color}
\usepackage{longtable,lscape}
\usepackage{multirow}
\usepackage{natbib}
\usepackage{xspace}
\bibpunct{(}{)}{;}{a}{}{,}

\newcommand{\kms}{km\,s$^{-1}$\xspace}
\newcommand{\mic}{$\mu$m\xspace}

\newcommand{\gastronoom}{\emph{GASTRoNOoM}\xspace}
\newcommand{\mcmax}{\emph{MCMax}\xspace}

\newcommand{\rstar}{R$_\star$\xspace}
\newcommand{\mg}{\dot{M}_\mathrm{g}}
\newcommand{\md}{\dot{M}_\mathrm{d}}

\newcommand{\msunyr}{\mathrm{M}_\odot\ \mathrm{yr}^{-1}}

\begin{document}


\title{Observational evidence for composite grains in an AGB outflow}
\subtitle{MgS in the extreme carbon star LL Peg}

\author{R.~Lombaert \inst{1}
\and B.L.~de Vries \inst{1}
\and A.~de Koter \inst{2,3}
\and L.~Decin \inst{1,2}
\and M.~Min \inst{2,3}
\and K.~Smolders \inst{1}
\and H.~Mutschke \inst{4}
\and L.B.F.M.~Waters \inst{2,5}
}


\institute{Instituut voor Sterrenkunde, KU Leuven, Celestijnenlaan 200D B-2401, 3001 Leuven, Belgium 
\and Astronomical Institute ``Anton Pannekoek'', University of Amsterdam, P.O.~Box 94249, 1090 GE Amsterdam, The Netherlands
\and Astronomical Institute Utrecht, University Utrecht, P.O.~Box 80000, 3508 TA Utrecht, The Netherlands
\and Astrophysikalisches Institut und Universit\"{a}ts-Sternwarte, Schillerg\"{a}\ss chen 2-3, 07745 Jena, Germany
\and Netherlands Institute for Space Research, Sorbonnelaan 2, 3584 CA Utrecht, The Netherlands
}

\date{Accepted 01 July 2012}

\authorrunning{R. Lombaert et al.}
\titlerunning{Observational evidence for composite grains in an AGB outflow}

\abstract {The broad 30 \mic feature in carbon stars is commonly attributed to MgS dust particles. However, reproducing the 30 \mic feature with homogeneous MgS grains would require much more sulfur relative to the solar abundance. Direct gas-phase condensation of MgS occurs at a low efficiency. Precipitation of MgS on SiC precursor grains provides a more efficient formation mechanism, such that the assumption of homogeneous MgS grains may not be correct. Using a Monte Carlo-based radiative transfer code, we aim to model the 30 \mic feature of the extreme carbon star LL Peg with MgS dust particles. We find that for LL Peg this modeling is insensitive to the unknown MgS optical properties at $\lambda <10 $ \mic. When MgS is allowed to be in thermal contact with amorphous carbon and SiC, the amount of MgS required to reproduce the strength of 30 \mic feature agrees with the solar abundance of sulfur, thereby resolving the reported \emph{MgS mass problem}. We conclude that MgS is a valid candidate to be the carrier of the 30 \mic feature when it is part of a composite grain population that has optical properties representative of an ensemble of particle shapes.}



\maketitle


\section{Introduction}\label{sect:intro}
\vspace{-0.1cm}
The broad 30 \mic feature in the thermal continuum emission of carbon stars was identified in 1985 by \citet{goe1985} as due to magnesium sulfide (MgS). \citet{beg1994} presented optical data for MgS and compared them to the thermal continuum emission of CW Leo to confirm MgS as the likely carrier of the 30 \mic feature. The \emph{Infrared Space Observatory} (ISO;  \citeauthor{kes1996}~\citeyear{kes1996}) observed a large sample of carbon stars (e.g.~\citeauthor{yam1998}~\citeyear{yam1998}; \citeauthor{jia1999}~\citeyear{jia1999}; \citeauthor{szc1999}~\citeyear{szc1999}; \citeauthor{hri2000}~\citeyear{hri2000}; \citeauthor{vol2002}~\citeyear{vol2002}) displaying a diversity in strength, shape and width of the feature.  
In an analysis of this data set, \citet{hon2002} concluded that the shape of the 30 \mic feature can be best reproduced when the grains are not perfect spheres. 
\citet{zha2009a} modeled the 30 \mic feature in the proto-planetary nebula HD 56126 using pure MgS grains. Assuming the grains to be irradiated by unattenuated stellar light, their analysis needed the optical properties of MgS at wavelengths $\lambda < 10$ \mic. Unfortunately, such measurements are lacking, as yet. Assuming relatively high absorption efficiencies in this regime, these authors required an amount of MgS up to ten times the amount of available atomic sulfur to explain the strength of the 30 \mic feature. Owing to this \emph{mass problem}, Zhang and collaborators argued against MgS as the carrier of the feature. 

%

In this letter, we report on the observational evidence for composite grains in the outflow of the high mass-loss rate asymptotic giant branch (AGB) star LL Peg, also known as AFGL 3068. We show that MgS is a viable candidate to explain the 30 \mic feature independent of the absorbing efficiencies of these grains at $\lambda < 10$ \mic. Moreover, we show that if one assumes thermal contact between the dust species in the outflow, the \emph{mass problem} in HD 56126 reported by \citet{zha2009a} does not occur in LL Peg. Finally, we discuss the formation of composite grains in AGB outflows.

\section{Data} \label{sect:data}
The spectral energy distribution (SED) of LL Peg was constructed by combining spectra taken with the \emph{Short Wavelength Spectrometer} (SWS) and \emph{Long Wavelength Spectrometer} (LWS) on board ISO. The SWS data were retrieved from the \citet{slo2003} database; the LWS data from the ISO data archive. The LWS data were rescaled to the calibrated SWS data. We corrected for interstellar reddening following the extinction law of \citet{chi2006}, with an extinction correction factor in the K-band of $A_\mathrm{K} = 0.01$ mag \citep{dri2003}. 
\section{Modeling the thermal energy distribution}\label{sect:contmodel}

\begin{figure}[!t]
\resizebox{\hsize}{!}{\includegraphics{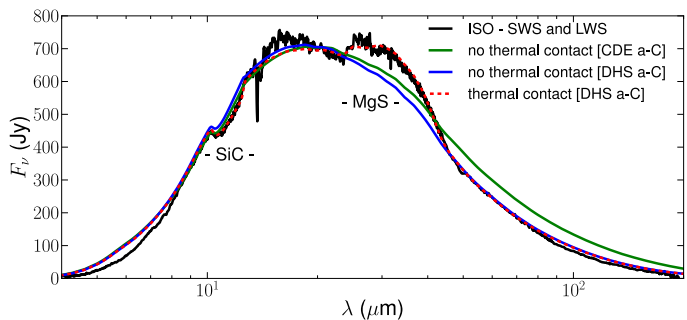}}
\vspace{-0.6cm}
\caption{SED of LL Peg. The SWS and LWS data are shown in full black. The dashed red curve shows a model assuming thermal contact between all dust species, whereas the full blue line shows a model assuming no thermal contact. The dust composition of both models consists of a-C grains in DHS shapes and SiC, MgS and Fe grains in CDE shapes. The full green model includes a-C grains in CDE shapes and assumes no thermal contact.}
\vspace{-0.1cm}
\label{fig:tc_notc}
\end{figure}
\begin{figure}[!t]
\resizebox{\hsize}{!}{\includegraphics{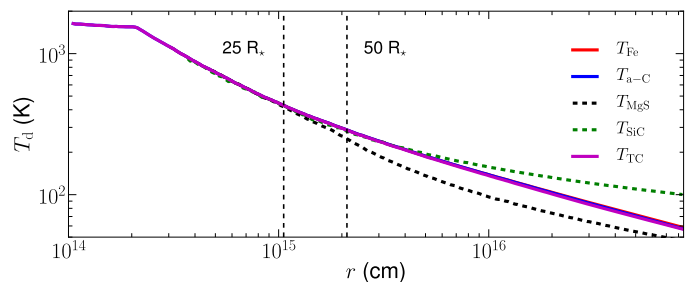}}
\vspace{-0.6cm}
\caption{Dust temperature profile of the envelope of LL Peg for individual species and composite grains: Fe in red, a-C in blue, MgS in dashed-black, SiC in dashed-green and composite grains in magenta. Note that the red, blue and magenta profiles essentially coincide throughout the whole envelope.}
\vspace{-0.4cm}
\label{fig:dusttemp}
\end{figure}

We combined two numerical codes to reproduce the ISO observations of LL Peg. The first is \mcmax \citep{min2009a}, a dust continuum radiative transfer code based on a Monte Carlo method \citep{bjo2001}, which computes the dust temperature self-consistently from the thermal energy balance equation. The second is \gastronoom \citep{dec2006,dec2010a}, which calculates the momentum transfer from dust to gas in the outflow. The results of these two modeling tools were iterated to achieve a self-consistent solution for the observed gas expansion velocity \citep{lom2012}. 
Recently, \citet{deb2010} found $\mg = 3.1 \times 10^{-5}\ \msunyr$ for LL Peg, but indicated a possible mass-loss variability. We assumed $\mg = 3 \times 10^{-5}\ \msunyr$, which leads to a dust expansion velocity of 16 \kms. The stellar effective temperature of LL Peg was taken to be $T_\star = 2400$ K. The adopted temperature does not impact the results because the CSE of LL Peg is optically thick up to $\lambda \sim 50$ \mic. We used a luminosity of $L_\star = 1.1 \times 10^{4}$ L$_\odot$ at a distance of $d_\star = 1300$ pc, following the period-luminosity relation of \citet{gro1996}. The stellar radius is thus $R_\star = 608$ R$_\odot$. 

Table~\ref{table:dust} lists the dust species used in this study and relevant dust properties. The bulk of the dust composition is provided by amorphous carbon (a-C). We included a small fraction of metallic iron (Fe), which is a common dust component in the outflow of carbon stars \citep{lat1978}. We used silicon carbide (SiC) to model the 11 \mic feature (see \citeauthor{spe2009}~\citeyear{spe2009}, and references therein for a more detailed discussion on SiC as a dust component in carbon-rich envelopes), and MgS to model the 30 \mic feature (see Sect.~\ref{sect:30model}). The optical properties of these dust species were derived from reflection measurements in the laboratory. References for these measurements are listed in Table~\ref{table:dust}. We used three models to represent grain shapes (\citeauthor{boh1983}~\citeyear{boh1983}; \citeauthor{min2003}~\citeyear{min2003}): a continuous distribution of ellipsoids (CDE), a distribution of hollow spheres with filling factor 0.8 (DHS), and spherical particles (MIE). We adopted a single grain size of 0.01 \mic for all grain shape models.

In constructing the optical properties of composite grains we summed up the extinction contributions of the separate dust species. This is a valid assumption if absorption and emission occur in the Rayleigh limit, i.e. at $\lambda \gg 0.01$ \mic \citep{min2008}. Moreover, the refractive indices of MgS are significantly higher in the 30 \mic feature than those of the other dust species included in the model and, for the relative abundances found in this study, dominate the spectral behavior of the composite grains in this wavelength region. This supports our approach. Alternatively, one may compute the optical properties of composite grains. This would require assuming a composite structure, which may add a larger uncertainty to the spectral shape than the assumption of summing the individual extinction contributions.

The 11 \mic feature is best fitted with SiC dust in a CDE ensemble. Using the CDE ensemble for a-C grains, however, significantly increases the emission at $\lambda > 50 $ \mic, as shown by the full green curve in Fig.~\ref{fig:tc_notc}. The full blue model, on the other hand, adopts the DHS particle shapes for a-C, yielding better results. We find a dust-mass-loss rate of $\md = (1.7 \pm 0.1) \times 10^{-7}\ \msunyr$, comprised of 70\% a-C, 5\% Fe, 10\% SiC and 15\% MgS. 

\section{The 30 \mic feature: resolving the mass problem}\label{sect:30model}
In principle, the result of a model including MgS may critically depend on the optical constants assumed for MgS at $\lambda < 10$ \mic, since the shape of the 30 \mic feature is very sensitive to the temperature distribution of the MgS grains \citep{hon2002}. For stars that have a low mass-loss rate this introduces a large uncertainty in the spectral behavior of MgS because the heating of this species is caused by direct stellar light at precisely these wavelengths. However, for a high mass-loss-rate star such as LL Peg, the model dependence on the unknown short wavelength MgS optical constants disappears. The wind of LL Peg is optically thick up to $\lambda \sim 50$ \mic with the transition from an optically thick to an optically thin envelope in the infrared occurring between 15 and 35 \rstar. For individual species the dust temperature profiles coincide in the optically thick region of the outflow, see Fig.~\ref{fig:dusttemp}. In the 
optically thin, outer wind the MgS and SiC profiles start to deviate from those of a-C and Fe. The 30 $\mu$m feature is produced by emission from dust particles in the outer envelope, which are heated by the infrared radiation emitted from an optically thick surface at $\sim 25$ \rstar. The dust temperature at this radial distance is $\sim 400$ K, such that most of the heating occurs at $\lambda > 7$ \mic. 
\begin{table}[!t]
	{
    	\setlength{\tabcolsep}{5pt}
    	\caption{Chemical formula, specific density $\rho_\mathrm{s}$, condensation temperature $T_\mathrm{cond}$, and condensation radius $R_\mathrm{cond}$ of the dust species. The final column lists the reference to the optical constants: 1.~\citet{jag1998b}; 2.~\citet{hen1996}; 3.~\citet{pit2008}; 4.~\citet{beg1994}.}\label{table:dust}
\vspace{-0.2cm}
  	\begin{center}
   	\begin{tabular}[c]{lrrrrr}\hline\hline\rule[0mm]{0mm}{3mm}Dust species	 &  Chem.    &   $\rho_\mathrm{s}$       & $T_\mathrm{cond}$   & $R_\mathrm{cond}$  &Ref. \\
                 &    form.             &   (g cm$^-3$)   & (K)      &     (R$_\star$)             &  \\\hline
   Amorphous carbon  &  a-C & 1.8 & 1650  &2.5& 1 \\
   Metallic iron &  Fe             &   7.9         & 1100        &     7.0          &  2 \\
   Silicon carbide     &  SiC         &     3.2      & 1400        &    4.0            & 3 \\
   Magnesium sulfide     & MgS         &   3.0         &  750      &     11.0            & 4 \\  	\hline
   	\end{tabular}
   	\end{center}
	}
\vspace{-0.6cm}
\end{table}

As indicated by \citet{zha2009a}, the strength of the 30 \mic feature in HD 56126 cannot be reproduced using pure MgS grains, unless the MgS abundance far exceeds the amount of atomic sulfur that is available from atmospheric abundance estimates. This \emph{mass problem} is shown in the full blue model in Fig.~\ref{fig:tc_notc}, which assumes an amount of MgS dust in agreement with the solar abundance of sulfur. However, \citeauthor{zha2009a} used chemically homogeneous grains, which heat and cool independently of grains of a different chemical composition. If all dust species are in thermal contact, the temperature distribution (shown by the magenta curve in Fig.~\ref{fig:dusttemp}) is predominantly set by the chemical component that is both reasonably abundant and is an efficient absorber. For a carbon-rich environment, a-C grains heat and cool more efficiently over a broader wavelength range than MgS or SiC grains. As shown by the dashed red curve in Fig.~\ref{fig:tc_notc}, assuming thermal contact between the dust species, but otherwise identical parameters as in the blue model, indeed results in a 30 \mic feature that is produced well in both strength and shape. 

We quantified our results by calculating the atomic number abundance with respect to H$_2$ of sulfur needed for the MgS dust mass in our models. We only considered sulfur, since it is the least abundant component with a solar atomic abundance of $(2.6 \pm 0.2) \times 10^{-5}$ \citep{asp2009}. Two models were calculated, one assuming thermal contact between dust species (see the dashed red curve in Fig.~\ref{fig:tc_notc}), and one assuming no thermal contact (see the dashed blue curve in Fig.~\ref{fig:mgsfits} for the continuum-divided 30 \mic feature). In both models, the dust composition and dust-mass-loss rate were adapted such that the 30 \mic feature is reproduced with a similar equivalent width. The model assuming thermal contact yields an atomic number abundance of $2.5 \times 10^{-5}$ for sulfur. The model without thermal contact requires an abundance of $1.2\times10^{-4}$. Given the potentially variable mass loss of LL Peg (e.g.~\citeauthor{mau2006}~\citeyear{mau2006}), we estimate the uncertainty on these values to be a factor of two. In the case of thermal contact, the required amount of MgS to reproduce the 30 \mic feature is within the limits imposed by a solar sulfur abundance, independent of the particle shape model. Without assuming thermal contact, the required amount of MgS would significantly exceed a solar atomic sulfur abundance. 

\section{Discussion} \label{sect:results}
\subsection{Homogeneous versus composite grains}
The above motivated need for thermal contact between the dust species implies that they must be included in some kind of heterogeneous composite grain structure. Recently, \citet{zhu2008} discussed the possibility of forming MgS dust in carbonaceous environments through precipitation on SiC precursor grains, which would be a plausible way to achieve composite grains. These authors showed that the formation of MgS is strongly coupled to that of SiC, a process in which sulfur is being freed by breaking up SiS molecules. This sulfur is stored in H$_2$S, which can react with freely available magnesium to condense into MgS. A connection between molecular SiS and the formation of MgS is also supported by \citet{smo2012}, who found a strong correlation between the presence of SiS molecular bands and a 30 \mic feature in their sample of S-stars. SiC is not formed in these stars, but SiS can react directly with Mg to form MgS, albeit less efficiently. This scenario agrees with the significantly lower MgS abundance that is needed to explain the typically weak 30 \mic feature in S-stars. 

Additional support for the above hypothesis was given by \citet{lei2008}, who compared galactic carbon stars with a sample in the LMC, in which all sources show an SiC feature at 11 \mic, while only half of the sources show emission at 30 \mic. Stronger 30 \mic features are found together with weaker SiC features, which led Leisenring and collaborators to suggest that MgS forms as a coating on top of SiC grains. In galactic AGB stars, a-C is expected to form first, followed by SiC. If it is energetically beneficial for SiC to form on top of a-C grains in a core-mantle structure, or for homogenous a-C and SiC grains to stick together in an aggregate structure, it is likely that composite grains are formed. For core-mantle grains, \citet{zhu2008} showed that a resonance effect caused by such a structure induces a second peak at $\sim 35$ \mic in the absorption efficiency profile of spherical MgS grains. However, an ensemble of non-spherical grains will likely not produce such a resonance effect, see Sect.~\ref{sect:shape}. 

MgS formation directly from the gas-phase -- without any kind of precursor grain of different composition -- would provide a mechanism to produce chemically homogeneous grains. However, several studies have indicated that this process is not sufficiently efficient to explain the large amounts of MgS needed to produce a strong 30 \mic feature \citep{kim2005,zhu2008,che2011b}. In their study of carbon-rich AGB stars and planetary nebulae, \citet{hon2002} adopted chemically homogeneous grains and derived a temperature for the MgS particles by fitting the 30 $\mu$m feature after subtraction of a smooth continuum. For LL Peg, they found a continuum (i.e.~essentially a-C) temperature of 340 K and a MgS temperature of 120 K. They ascribed the difference to the chemical homogeneity of the grains. However, the authors assumed that MgS is formed in an optically thin medium and can be characterized by a unique temperature. Our radiative transfer calculations show that both these assumptions are not valid for LL Peg. Moreover, the low temperature of MgS found by \citeauthor{hon2002} leads to the MgS \emph{mass problem}. We conclude that a population of chemically homogeneous grains, and hence no thermal contact between dust species, cannot be reconciled with the spectrum of LL Peg.
\subsection{Particle shape and size}\label{sect:shape}
\begin{figure}[!t]
\resizebox{\hsize}{!}{\includegraphics{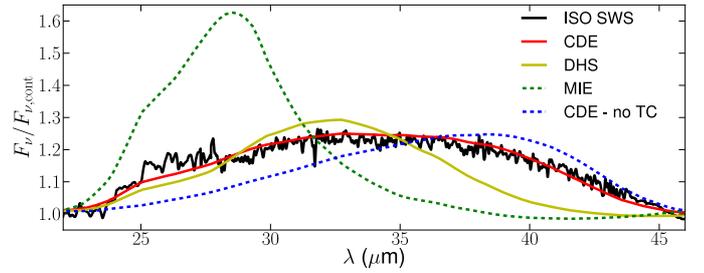}}
\vspace{-0.6cm}
\caption{Continuum-divided 30 \mic feature with the SWS data of LL Peg shown in full black. The CDE model in full red assumes thermal contact and requires a sulfur abundance of about the solar value. In the model represented by the dashed blue line the CDE particles are not in thermal contact. To fit the strength of the feature sulfur needs to be $\sim 5$ times the solar value. A DHS (full yellow line) and MIE (dashed green) model represent alternative shape distributions of particles in which the dust components are in thermal contact. The sulfur abundance in the MIE model has been adjusted to match the equivalent width of the observed feature.}
\vspace{-0.5cm}
\label{fig:mgsfits}
\end{figure}
The spectral shape of a dust emission feature may strongly depend on the model that is used to describe the particle shapes \citep{hon2002,mut2009}. The continuum-divided MgS features for several particle models are shown in Fig.~\ref{fig:mgsfits}. Mie particles clearly cannot reproduce the spectral profile around 30 \mic in LL Peg. Several studies (e.g.~\citeauthor{hon2002}~\citeyear{hon2002}, \citeauthor{min2003}~\citeyear{min2003}) have indicated that Mie particles in any grain-size distribution cause narrow features and certain resonances, which have not been observed in the thermal continuum emission of AGB stars. It is generally better to use an ensemble of particle shapes, such as given by a CDE or DHS (see \citeauthor{mut2009}~\citeyear{mut2009} for an overview of these and more advanced particle shape models), because they more accurately represent features caused by a collection of irregularly shaped dust particles. Note that for both the CDE and the DHS model the extinction properties are independent of our single grain-size assumption as long as the Rayleigh limit holds. The assumption of homogeneous or composite grains also impacts the shape of the MgS feature, because the dust temperature profile is affected. If one assumes homogeneous grains (i.e.~no thermal contact) and increases the MgS-dust mass such that the strength of the feature is approximated, one finds that the 30 \mic feature shape is rather poorly reproduced by a CDE model (see the dashed blue curve in Fig.~\ref{fig:mgsfits}). However, this poor agreement cannot be used as an argument for the need of thermal contact because CDE is quite a simple grain shape model. A comparative study between particle shape models by \citet{mut2009} has shown that the CDE model often results in emission bands too much enhanced at red wavelengths when compared to dust optical properties measured in the laboratory. The good fit of our MgS CDE model given by the red curve and the observed 30 \mic feature in LL Peg consequently may indicate some additional effect that enhances the red wing of the observed band profile. Condensation experiments by \citet{kim2005} have shown that MgS dust formed as network-like structures can show an enhanced red part of the MgS band profile. Spectra of dust grains embedded in a matrix also show such an enhancement compared to those of free particles (e.g. \citeauthor{tam2006}~\citeyear{tam2006}), which may point to mixing or embedding of the MgS dust in inhomogeneous grains.
\begin{figure}[!t]
\begin{center}$
\begin{array}{cc}
\resizebox{8.4cm}{!}{\includegraphics{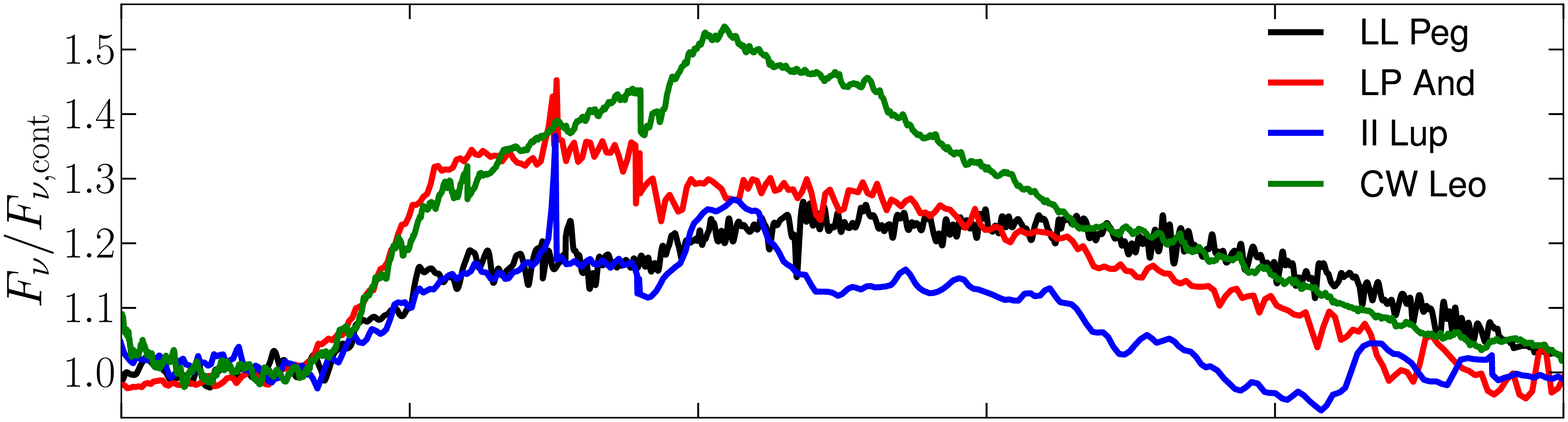}} \\
\resizebox{8.4cm}{!}{\includegraphics{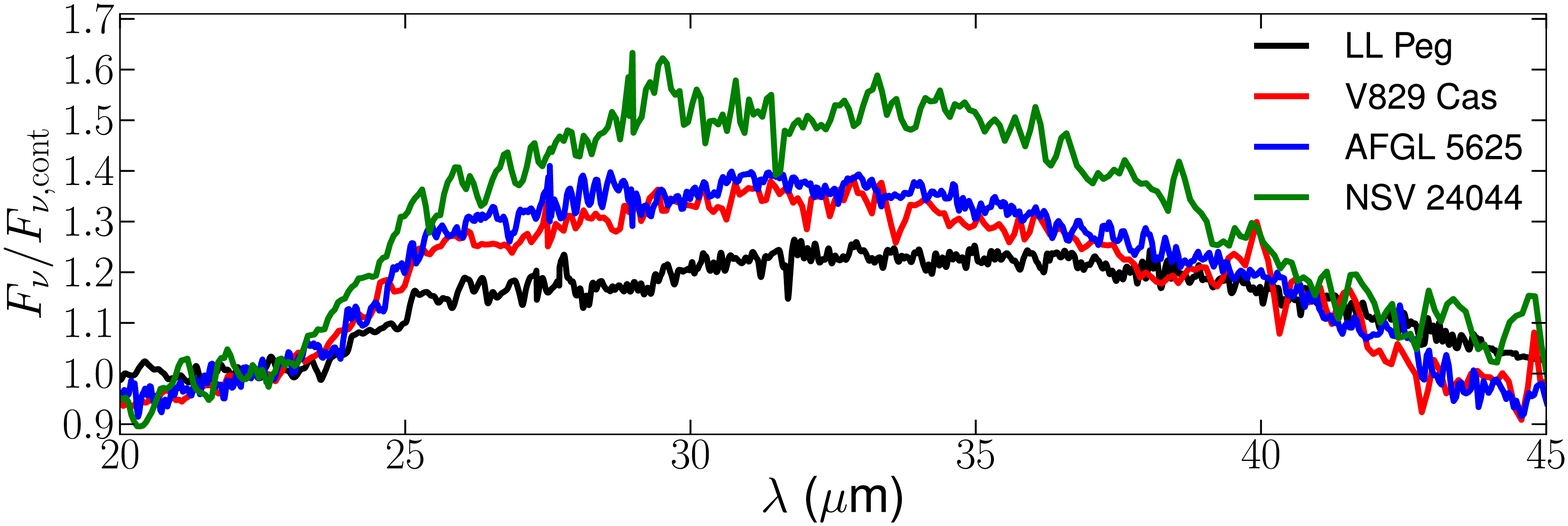}} 
\end{array}$
\end{center}
\vspace{-0.6cm}
\caption{Continuum-divided ISO spectra of multiple carbon stars in the region of the 30 \mic feature, with LL Peg as a reference in black. The top panel shows three typical AGB stars with relatively low-density envelopes; the bottom panel shows three typical AGB stars with relatively high density-envelopes (data taken from \citeauthor{hon2002}~\citeyear{hon2002}).}
\label{fig:allstars}
\vspace{-0.5cm}
\end{figure}
\subsection{Diversity of the 30 \mic feature shape in AGB outflows}
LL Peg is a highly evolved AGB star and exhibits one of the reddest thermal emission continua of all carbon stars. The outflow of this source is optically thick even in the 30 \mic feature. As indicated in the bottom panel of Fig.~\ref{fig:allstars}, AGB stars with a high-density envelope show an almost uniform shape of the 30 \mic feature. Sources with a lower density envelope, such as those shown in the top panel of Fig.~\ref{fig:allstars}, display a wide diversity in shape. As a first possible explanation of this phenomenon, \citet{kim2005} suggested that the shape of the 30 \mic feature may depend on the formation history of MgS. If the grains have formed through gas-phase condensation, the resulting emission feature is expected to be more pronounced at $\lambda \sim 32$ \mic than at longer wavelengths. If MgS grains are produced via a gas-solid reaction, the red wing of the feature at $\lambda > 35$ \mic becomes more intense relative to the strength at $\lambda \sim 32$ \mic. As shown in Fig.~\ref{fig:allstars}, the 30 \mic feature in the high-density envelopes has a more pronounced red wing than those in the lower density envelopes. One may speculate that in low-density sources, where the 30 \mic feature is optically thin throughout the entire envelope, the feature may be composed of gas-phase condensates in the inner part of the wind and grains formed by gas-solid reactions throughout or in the outer part of the wind. In high-density sources, one only observes the contributions from the latter formation channel. This scenario could be consistent with the MgS-SiC correlation in S-type stars \citep{smo2012}, where the lack of SiC particles indicates a gas-phase condensation channel, albeit not a very effective one. A second explanation might be that the low opacities in low-density envelopes favor the heating of another dust species with an absorption coefficient profile peaking at about 30 \mic. Alternative carriers suggested in the literature include hydrogenated amorphous carbons \citep{gri2001}.

\section{Conclusions}\label{sect:conclusion}
We have modeled the ISO spectrum of the high-density carbon star LL Peg with a dust composition consisting of a-C, Fe, SiC, and MgS. The (high) density and temperature structure in the envelope of this source allow one to model the 30 \mic feature with MgS dust particles, independent of the unknown MgS optical properties at $\lambda < 10$ \mic. We showed that MgS is a viable candidate to be the carrier of the 30 \mic feature. An ensemble of particle shapes works significantly better than spherical grains to explain the shape of the feature. Thermal contact between the dust species is required to ensure that the amount of MgS dust in the envelope of LL Peg does not exceed the solar abundance of sulfur, thereby avoiding the \emph{mass problem} as reported by \citet{zha2009a} for the post-AGB star HD 56126. Achieving thermal contact between all dust species is possible if these species form in some kind of heterogeneous composite grain structure. 
\begin{acknowledgements}
We thank S.~Hony for providing the reduced SWS spectra of the high-density carbon stars. We also express gratitude toward the referee, who provided instructive feedback. RL acknowledges financial support from the Fund for Scientific Research - Flanders (FWO) under grant number ZKB5757-04-W01, and from the Department of Physics and Astronomy of the KULeuven. BdV and KS acknowledge financial support from the FWO.
\end{acknowledgements}

\bibliographystyle{aa}
\bibliography{allreferences}

\begin{thebibliography}{38}
\expandafter\ifx\csname natexlab\endcsname\relax\def\natexlab#1{#1}\fi

\bibitem[{{Asplund} {et~al.}(2009){Asplund}, {Grevesse}, {Sauval}, \&
  {Scott}}]{asp2009}
{Asplund}, M., {Grevesse}, N., {Sauval}, A.~J., \& {Scott}, P. 2009, \araa, 47,
  481

\bibitem[{{Begemann} {et~al.}(1994){Begemann}, {Dorschner}, {Henning},
  {Mutschke}, \& {Thamm}}]{beg1994}
{Begemann}, B., {Dorschner}, J., {Henning}, T., {Mutschke}, H., \& {Thamm}, E.
  1994, \apjl, 423, L71

\bibitem[{{Bjorkman} \& {Wood}(2001)}]{bjo2001}
{Bjorkman}, J.~E. \& {Wood}, K. 2001, \apj, 554, 615

\bibitem[{{Bohren} \& {Huffman}(1983)}]{boh1983}
{Bohren}, C.~F. \& {Huffman}, D.~R. 1983, {Absorption and scattering of light
  by small particles}, ed. C.~F. {Bohren} \& D.~R. {Huffman}

\bibitem[{{Cherchneff}(2011)}]{che2011b}
{Cherchneff}, I. 2011, ArXiv e-prints

\bibitem[{{Chiar} \& {Tielens}(2006)}]{chi2006}
{Chiar}, J.~E. \& {Tielens}, A.~G.~G.~M. 2006, \apj, 637, 774

\bibitem[{{De Beck} {et~al.}(2010){De Beck}, {Decin}, {de Koter}, {Justtanont},
  {Verhoelst}, {Kemper}, \& {Menten}}]{deb2010}
{De Beck}, E., {Decin}, L., {de Koter}, A., {et~al.} 2010, \aap, 523, A18

\bibitem[{{Decin} {et~al.}(2010){Decin}, {De Beck}, {Br{\"u}nken},
  {M{\"u}ller}, {Menten}, {Kim}, {Willacy}, {de Koter}, \&
  {Wyrowski}}]{dec2010a}
{Decin}, L., {De Beck}, E., {Br{\"u}nken}, S., {et~al.} 2010, \aap, 516, A69

\bibitem[{{Decin} {et~al.}(2006){Decin}, {Hony}, {de Koter}, {Justtanont},
  {Tielens}, \& {Waters}}]{dec2006}
{Decin}, L., {Hony}, S., {de Koter}, A., {et~al.} 2006, \aap, 456, 549

\bibitem[{{Drimmel} {et~al.}(2003){Drimmel}, {Cabrera-Lavers}, \&
  {L{\'o}pez-Corredoira}}]{dri2003}
{Drimmel}, R., {Cabrera-Lavers}, A., \& {L{\'o}pez-Corredoira}, M. 2003, \aap,
  409, 205

\bibitem[{{Goebel} \& {Moseley}(1985)}]{goe1985}
{Goebel}, J.~H. \& {Moseley}, S.~H. 1985, \apjl, 290, L35

\bibitem[{{Grishko} {et~al.}(2001){Grishko}, {Tereszchuk}, {Duley}, \&
  {Bernath}}]{gri2001}
{Grishko}, V.~I., {Tereszchuk}, K., {Duley}, W.~W., \& {Bernath}, P. 2001,
  \apjl, 558, L129

\bibitem[{{Groenewegen} \& {Whitelock}(1996)}]{gro1996}
{Groenewegen}, M.~A.~T. \& {Whitelock}, P.~A. 1996, \mnras, 281, 1347

\bibitem[{{Henning} \& {Stognienko}(1996)}]{hen1996}
{Henning}, T. \& {Stognienko}, R. 1996, \aap, 311, 291

\bibitem[{{Hony} {et~al.}(2002){Hony}, {Waters}, \& {Tielens}}]{hon2002}
{Hony}, S., {Waters}, L.~B.~F.~M., \& {Tielens}, A.~G.~G.~M. 2002, \aap, 390,
  533

\bibitem[{{Hrivnak} {et~al.}(2000){Hrivnak}, {Volk}, \& {Kwok}}]{hri2000}
{Hrivnak}, B.~J., {Volk}, K., \& {Kwok}, S. 2000, \apj, 535, 275

\bibitem[{{J{\"a}ger} {et~al.}(1998){J{\"a}ger}, {Mutschke}, \&
  {Henning}}]{jag1998b}
{J{\"a}ger}, C., {Mutschke}, H., \& {Henning}, T. 1998, \aap, 332, 291

\bibitem[{{Jiang} {et~al.}(1999){Jiang}, {Szczerba}, \& {Deguchi}}]{jia1999}
{Jiang}, B.~W., {Szczerba}, R., \& {Deguchi}, S. 1999, \aap, 344, 918

\bibitem[{{Kessler} {et~al.}(1996){Kessler}, {Steinz}, {Anderegg}, {Clavel},
  {Drechsel}, {Estaria}, {Faelker}, {Riedinger}, {Robson}, {Taylor}, \&
  {Xim{\'e}nez de Ferr{\'a}n}}]{kes1996}
{Kessler}, M.~F., {Steinz}, J.~A., {Anderegg}, M.~E., {et~al.} 1996, \aap, 315,
  L27

\bibitem[{{Kimura} {et~al.}(2005){Kimura}, {Kurumada}, {Tamura}, {Koike},
  {Chihara}, \& {Kaito}}]{kim2005}
{Kimura}, Y., {Kurumada}, M., {Tamura}, K., {et~al.} 2005, \aap, 442, 507

\bibitem[{{Lattimer} {et~al.}(1978){Lattimer}, {Schramm}, \&
  {Grossman}}]{lat1978}
{Lattimer}, J.~M., {Schramm}, D.~N., \& {Grossman}, L. 1978, \apj, 219, 230

\bibitem[{{Leisenring} {et~al.}(2008){Leisenring}, {Kemper}, \&
  {Sloan}}]{lei2008}
{Leisenring}, J.~M., {Kemper}, F., \& {Sloan}, G.~C. 2008, \apj, 681, 1557

\bibitem[{{Lombaert} {et~al.}(2012){Lombaert}, {Decin}, {de Koter},
  {Blommaert}, {Royer}, {De Beck}, {de Vries}, {Khouri}, \& {Min}}]{lom2012}
{Lombaert}, R., {Decin}, L., {de Koter}, A., {et~al.} 2012, \aap, submitted

\bibitem[{{Mauron} \& {Huggins}(2006)}]{mau2006}
{Mauron}, N. \& {Huggins}, P.~J. 2006, \aap, 452, 257

\bibitem[{{Min} {et~al.}(2009){Min}, {Dullemond}, {Dominik}, {de Koter}, \&
  {Hovenier}}]{min2009a}
{Min}, M., {Dullemond}, C.~P., {Dominik}, C., {de Koter}, A., \& {Hovenier},
  J.~W. 2009, \aap, 497, 155

\bibitem[{{Min} {et~al.}(2003){Min}, {Hovenier}, \& {de Koter}}]{min2003}
{Min}, M., {Hovenier}, J.~W., \& {de Koter}, A. 2003, \aap, 404, 35

\bibitem[{{Min} {et~al.}(2008){Min}, {Hovenier}, {Waters}, \& {de
  Koter}}]{min2008}
{Min}, M., {Hovenier}, J.~W., {Waters}, L.~B.~F.~M., \& {de Koter}, A. 2008,
  \aap, 489, 135

\bibitem[{{Mutschke} {et~al.}(2009){Mutschke}, {Min}, \& {Tamanai}}]{mut2009}
{Mutschke}, H., {Min}, M., \& {Tamanai}, A. 2009, \aap, 504, 875

\bibitem[{{Pitman} {et~al.}(2008){Pitman}, {Hofmeister}, {Corman}, \&
  {Speck}}]{pit2008}
{Pitman}, K.~M., {Hofmeister}, A.~M., {Corman}, A.~B., \& {Speck}, A.~K. 2008,
  \aap, 483, 661

\bibitem[{{Sloan} {et~al.}(2003){Sloan}, {Kraemer}, {Price}, \&
  {Shipman}}]{slo2003}
{Sloan}, G.~C., {Kraemer}, K.~E., {Price}, S.~D., \& {Shipman}, R.~F. 2003,
  \apjs, 147, 379

\bibitem[{{Smolders} {et~al.}(2012){Smolders}, {Neyskens}, {Blommaert}, {Hony},
  {Van Winckel}, {Decin}, {Van Eck}, {Sloan}, {Cami}, {Uttenthaler},
  {Degroote}, {Barry}, {Feast}, {Groenewegen}, {Matsuura}, {Menzies}, {Sahai},
  {van Loon}, {Zijlstra}, {Acke}, {Bloemen}, {Cox}, {de Cat}, {Desmet},
  {Exter}, {Ladjal}, {Ostensen}, {Saesen}, {van Wyk}, {Verhoelst}, \&
  {Zima}}]{smo2012}
{Smolders}, K., {Neyskens}, P., {Blommaert}, J.~A.~D.~L., {et~al.} 2012, ArXiv
  e-prints

\bibitem[{{Speck} {et~al.}(2009){Speck}, {Corman}, {Wakeman}, {Wheeler}, \&
  {Thompson}}]{spe2009}
{Speck}, A.~K., {Corman}, A.~B., {Wakeman}, K., {Wheeler}, C.~H., \&
  {Thompson}, G. 2009, \apj, 691, 1202

\bibitem[{{Szczerba} {et~al.}(1999){Szczerba}, {Henning}, {Volk}, {Kwok}, \&
  {Cox}}]{szc1999}
{Szczerba}, R., {Henning}, T., {Volk}, K., {Kwok}, S., \& {Cox}, P. 1999, \aap,
  345, L39

\bibitem[{{Tamanai} {et~al.}(2006){Tamanai}, {Mutschke}, {Blum}, \&
  {Meeus}}]{tam2006}
{Tamanai}, A., {Mutschke}, H., {Blum}, J., \& {Meeus}, G. 2006, \apjl, 648,
  L147

\bibitem[{{Volk} {et~al.}(2002){Volk}, {Kwok}, {Hrivnak}, \&
  {Szczerba}}]{vol2002}
{Volk}, K., {Kwok}, S., {Hrivnak}, B.~J., \& {Szczerba}, R. 2002, \apj, 567,
  412

\bibitem[{{Yamamura} {et~al.}(1998){Yamamura}, {de Jong}, {Justtanont}, {Cami},
  \& {Waters}}]{yam1998}
{Yamamura}, I., {de Jong}, T., {Justtanont}, K., {Cami}, J., \& {Waters}, L.
  1998, \apss, 255, 351

\bibitem[{{Zhang} {et~al.}(2009){Zhang}, {Jiang}, \& {Li}}]{zha2009a}
{Zhang}, K., {Jiang}, B.~W., \& {Li}, A. 2009, \apj, 702, 680

\bibitem[{{Zhukovska} \& {Gail}(2008)}]{zhu2008}
{Zhukovska}, S. \& {Gail}, H.-P. 2008, \aap, 486, 229

\end{thebibliography}
\end{document}